\documentclass[prl,twocolumn,preprintnumbers,superscriptaddress]{revtex4-2}
\usepackage{amssymb}
\usepackage{amsmath}
\usepackage{amsfonts}
\usepackage{graphicx}
\usepackage{dcolumn}
\usepackage{bm}
\usepackage{verbatim}
\usepackage{color}
\usepackage{wasysym}
\usepackage{xcolor}
\usepackage{ulem}
\usepackage{appendix}
\usepackage{MnSymbol}
\usepackage{tipa}

\begin{document}

\title{Supersolid Phases in Ultracold Gases of Microwave Shielded Polar Molecules}
\author{Wei Zhang}
\thanks{These authors contributed equally to this work.}
\affiliation{Institute of Theoretical Physics, Chinese Academy of Sciences, Beijing 100190, China} 
\affiliation{School of Physical Sciences, University of Chinese Academy of Sciences, Beijing 100049, China}

\author{Hongye Liu}
\thanks{These authors contributed equally to this work.}
\affiliation{Institute of Theoretical Physics, Chinese Academy of Sciences, Beijing 100190, China} 
\affiliation{School of Physical Sciences, University of Chinese Academy of Sciences, Beijing 100049, China}

\author{Fulin Deng}
\affiliation{Institute of Theoretical Physics, Chinese Academy of Sciences, Beijing 100190, China} 

\author{Kun Chen}
\email{chenkun@itp.ac.cn}
\affiliation{Institute of Theoretical Physics, Chinese Academy of Sciences, Beijing 100190, China} 
\affiliation{School of Physical Sciences, University of Chinese Academy of Sciences, Beijing 100049, China}

\author{Su Yi}
\email{yisu@nbu.edu.cn}
\affiliation{Institute of Fundamental Physics and Quantum Technology \& School of Physics Science and Technology, Ningbo University, Ningbo, 315211, China}

\author{Tao Shi}
\email{tshi@itp.ac.cn}
\affiliation{Institute of Theoretical Physics, Chinese Academy of Sciences, Beijing 100190, China} 
\affiliation{School of Physical Sciences, University of Chinese Academy of Sciences, Beijing 100049, China}

\date{\today }

\begin{abstract}
We propose a novel scheme to realize the supersolid phase in ultracold gases of microwave-shielded polar molecules by engineering an additional anisotropy in inter-molecular dipolar interaction via an elliptically polarized microwave. It is shown through quantum Monte-Carlo calculations that the interplay of the anisotropies between the interaction and trapping potential gives rise to rich quantum phases. Particularly, it is found that the supersolid phase emerges in the parameter regime accessible to current experiments. Our study paves the way for exploring the properties of supersolid phases in ultracold gases of polar molecules.
\end{abstract}

\maketitle

\textit{Introduction}.---Recent realization of ultracold gases of microwave-shielded polar molecules (MSPMs)~\cite{Luo2022a,Luo2022b,chen2023,Wang2023,Will2023b} provides a powerful new platform for exploring correlated many-body phases with unprecedented control. On the one hand, the realization of degenerate NaK molecular gases~\cite{Luo2022a,Luo2022b,chen2023} and, subsequently, the ultracold tetramers~\cite{chen2023} open up a new path toward the creation of the long-sought $p$-wave superfluid. On the other hand, the realization of NaCs condensate~\cite{Will2023b} bridges the gap between weakly interacting atomic gases and strongly interacting helium-4~\cite{Jin2024,Langen2024,Zhang2025}, which paves the way for realizing the elusive supersolid (SS) phase~\cite{Gross1957,Andreev1969,Leggett1970,Chester1970} beyond the helium paradigm~\cite{theo2012,exp2012}. 

Supersolidity---marked by the coexistence of crystalline order and superfluidity---has attracted extensive interest in ultracold atomic systems, including spin-orbit-coupled Bose-Einstein condensates~\cite{supersolidSO2017}, atoms in optical cavities~\cite{supersolidcavity2017}, and magnetic dipolar gases~\cite{supersolid2019,metasupersolid2019,supersolidPfau2019,supersolidFerlaino2019}. Fundamentally different from atoms, MSPMs exhibit a competition between long-range dipole-dipole interactions (DDI) and a strong short-range repulsive core due to shielding. This leads to antibunched density-density correlations and a significantly reduced condensate fraction~\cite{Jin2024,Zhang2025}, placing the system beyond the scope of the Gross-Pitaevskii equation and its Lee-Huang-Yang corrections~\cite{Langen2024}. Although one might expect that increasing interaction strength and density could drive the system into a SS phase, two key limitations arise: i) Experimentally, enhancing DDI~\cite{Will2023b,Deng2025,Karman2025} inevitably reduces the shielding effect, leading to substantial three-body losses~\cite{Wang2023}. ii) Theoretically, a strong attractive DDI induces short interparticle distances, where the repulsive shielding potential suppresses long-range phase coherence, drastically reducing the condensate fraction~\cite{Jin2024,Zhang2025}. A striking example is the emergence of a monolayer crystal under strong DDI with single-microwave shielding, where the superfluid fraction vanishes~\cite{Pohl2025}. Therefore, an experimentally feasible scheme for the SS phase should ensure that molecular gases remain at relatively low density.

In this Letter, we propose a scheme to realize the SS phase in ultracold gases of polar molecules shielded by dual microwaves. To this end, we first show that, by replacing the $\sigma$-polarized microwave with an elliptically polarized one, the inter-molecular dipolar interaction breaks its cylindrical symmetry around the $z$ axis. We derive the analytic expression for the effective interaction potential, in which the ellipticity of the microwave acts as a control knob for tuning the anisotropy of the dipolar interaction in the $xy$ plane. We then investigate the quantum phases of the trapped NaCs gases using the unbiased path-integral Monte Carlo (PIMC) simulations combined with the worm algorithm (WA)~\cite{prokof1998exact,Boninsegni2006a,Boninsegni2006b,chen2014universal,chen2013deconfined,he4,kuklov2024transverse,plasma2023}. On the parameter plane formed by the ellipticity and trap geometry, we map out the finite-temperature phase diagram which consists of the expanding gases (EG), SS, and self-bound droplet (SBD) phases. In particular, the SS phase appears in the parameter regime accessible to the current experiment. More importantly, because the emergence of the SS phase in our scheme takes full advantage of anisotropy of the dipolar interaction, the density of gas required for density modulation is much lower than that without ellipticity, which leads to a significantly extended lifetime of molecular gases.

\begin{figure}[tbp]
\includegraphics[trim=50 20 50 10, clip,width=1\linewidth]{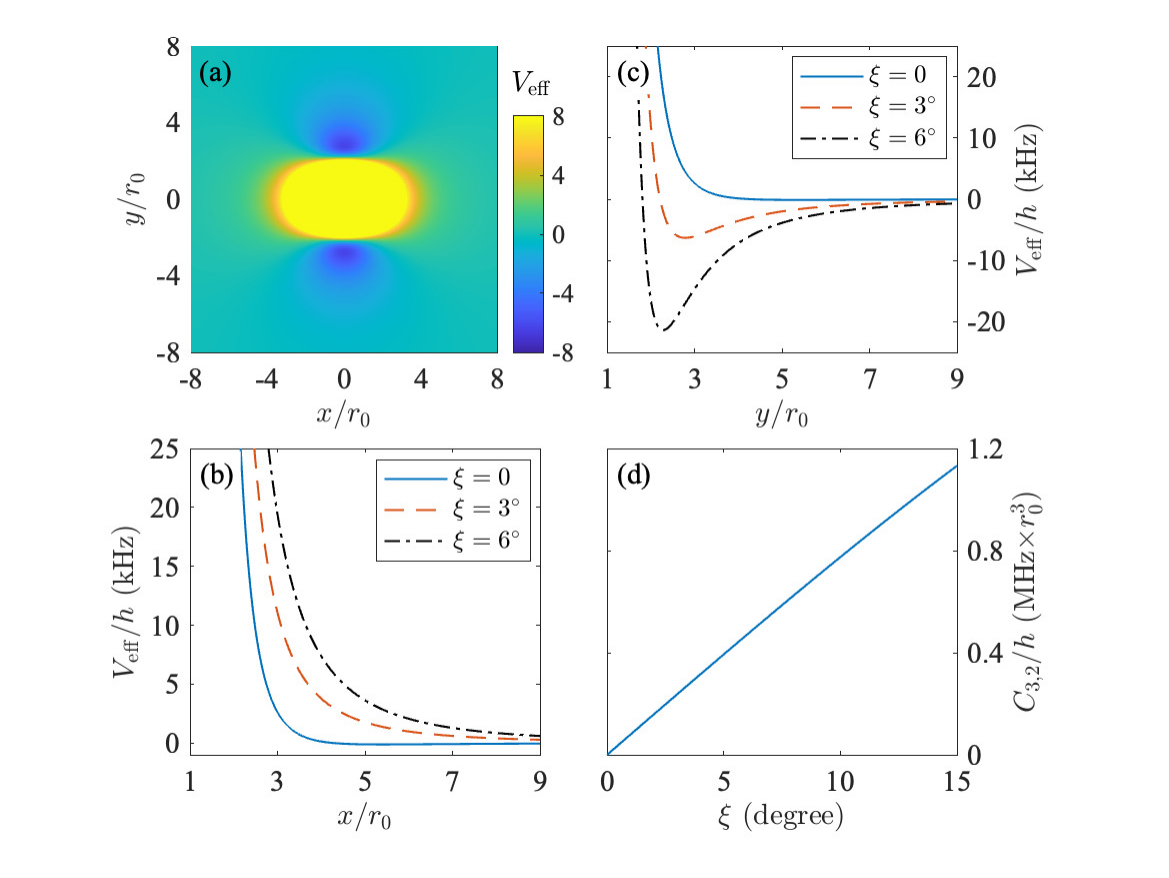}
\caption{(a) Effective potential (in units of kHz) on the $xy$ plane for $\xi=3^\circ$. (b) and (c) show the effective potentials along the $x$ and $y$ axes, respectively, for various values of $\xi$. (d) $C_{3,2}$ as a function of $\xi$.}
\label{figeffpot}
\end{figure}

\textit{Effective inter-molecular interaction}.---Before exploring the many-body properties, let us first determine the effective potential between two molecules shielded by a $\pi$-polarized and an elliptically polarized (with ellipticity $\xi$) microwaves. In this configuration, the amplitudes of the $\sigma^+$- and $\sigma^-$-polarized microwave fields are proportional to, $\cos\xi$ and $\sin\xi$, respectively. Then for $|\xi|\apprle 15^\circ$, the effective potential, up to the second-order perturbation, takes the form~\cite{SM}
\begin{align}
V_{\mathrm{eff}}({\boldsymbol{r}}) &=\frac{1}{r^{3}}\left[C_{3,0}(3\cos
^{2}\theta -1)+C_{3,2}\sin ^{2}\theta \cos 2\varphi \right]  \notag \\
&\quad+\frac{1}{r^{6}}\sum_{m,m^{\prime}=-2}^2C_{6,mm^{\prime }}Y_{2m}^{\ast
}(\hat{{\boldsymbol{r}}})Y_{2m^{\prime }}(\hat{{\boldsymbol{r}}}),\label{effpot}
\end{align}
where $\theta $ and $\varphi $ are, respectively, the polar and azimuthal angles of ${\boldsymbol r}$. As can be seen, the $1/r^3$ dipolar interaction consists of an isotropic and an anisotropic (on the $xy$ plane) parts, characterized by $C_{3,0}$ and $C_{3,2}$, respectively. Moreover, the $1/r^6$ shielding potential is governed by a $5\times 5$ symmetric matrix $C_6$ for interaction strengths. The parity symmetry of the potential further requires $C_{6,mm}=C_{6,-m,-m}$ and $C_{6,mm'}=0$ for $|m-m^{\prime}|\neq0,2,4$. Consequently, the shielding potential is completely determined by six independent parameters, say $C_{6,00}$, $C_{6,11}$, $C_{6,1-1}$, $C_{6,22}$, $C_{6,20}$, and $C_{6,2-2}$. As shown in the Supplemental Material (SM)~\cite{SM}, all interaction parameters are completely determined by the permanent dipole moment $d$ of the molecules, the ellipticity $\xi$, along with Rabi frequencies ($\Omega_\sigma, \Omega_\pi$) and detunings ($\delta_\sigma, \delta_\pi$) of the $\sigma$- and $\pi$-polarized microwave fields. Moreover, it can be verified that $V_{\rm eff}$ regains its cylindrical symmetry when $\xi=0$. The validity of these analytic results can be justified by directly comparing them with the adiabatic potential~\cite{SM}. In general, $C_3$'s are exact since they describe the long-range interaction; the second-order perturbation results for $C_{6}$'s are fairly accurate, as shown in SM. Nevertheless, the values of $C_6$'s can be improved by numerically fitting Eq.~\eqref{effpot} with the adiabatic potential. Further justification of the effective potential can be performed by comparing the scattering properties calculated using the effective potential with those by multi-channel calculations~\cite{SM}.

To visualize the effective potential, we consider, as a concrete example, the NaCs molecules with a dipole moment $d=4.6\,{\rm Debye}$. Following the experimental setup in Ref.~\cite{Will2023b}, the microwave parameters are chosen as $\Omega_{\sigma}=2\pi \times 7.9\,\mathrm{MHz}$, $\delta_{\sigma}=-2\pi \times 8\,\mathrm{MHz}$, and $\delta_{\pi}=-2\pi \times 10\,\mathrm{MHz}$. In addition, we assume that $\Omega_\pi=2\pi\times 6.5\,{\rm MHz}$, which leads to $C_{3,0}\approx0$. As a result, the $1/r^3$ dipolar interaction is solely contributed by the anisotopic part $C_{3,2}$. In Fig.~\ref{figeffpot}(a), we map out the typical effective potential in the $xy$ plane with $\xi=3^\circ$. The long-range behavior of the potential can be understood by noting that $C_{3,2}$ is proportional to $\sin2\xi$~\cite{SM} and its sign is uniquely determined by $\xi$. Consequently, the dipolar interaction is repulsive (attractive) along the $x$ ($y$) axis. In addition, the long-range dipolar interaction along the $z$ direction nearly vanishes. As to the shielding potential, the size of the shielding potential is over $r_0$ ($\equiv 10^3a_0$ with $a_0$ being the Bohr radii). It remains repulsive along all three dimensions even for nonzero $\xi$. Since changing the sign of $\xi$ only rotates the potential along the $z$ axis by $\pi/2$, we assume, without loss of generality, that $\xi\geq0$ throughout this work. To reveal more details about the effective potential, we plot, in Fig.~\ref{figeffpot}(b) and (c), the $x$ and $y$ dependence of the potential, respectively, for various $\xi$'s. Since the repulsion (attraction) along the $x$ ($y$) axis is strengthened with $\xi$, the planar anisotropy is enhanced by increasing $\xi$. This can also be seen in Fig.~\ref{figeffpot}(d) where $C_{3,2}$ appears as an increasing function of $\xi$. Finally, we point out that the effective potential along the $z$ axis remains roughly unchanged with $\xi$.


\begin{figure}[tbp]
\includegraphics[trim=20 210 20 220, clip,width=1.0\linewidth]{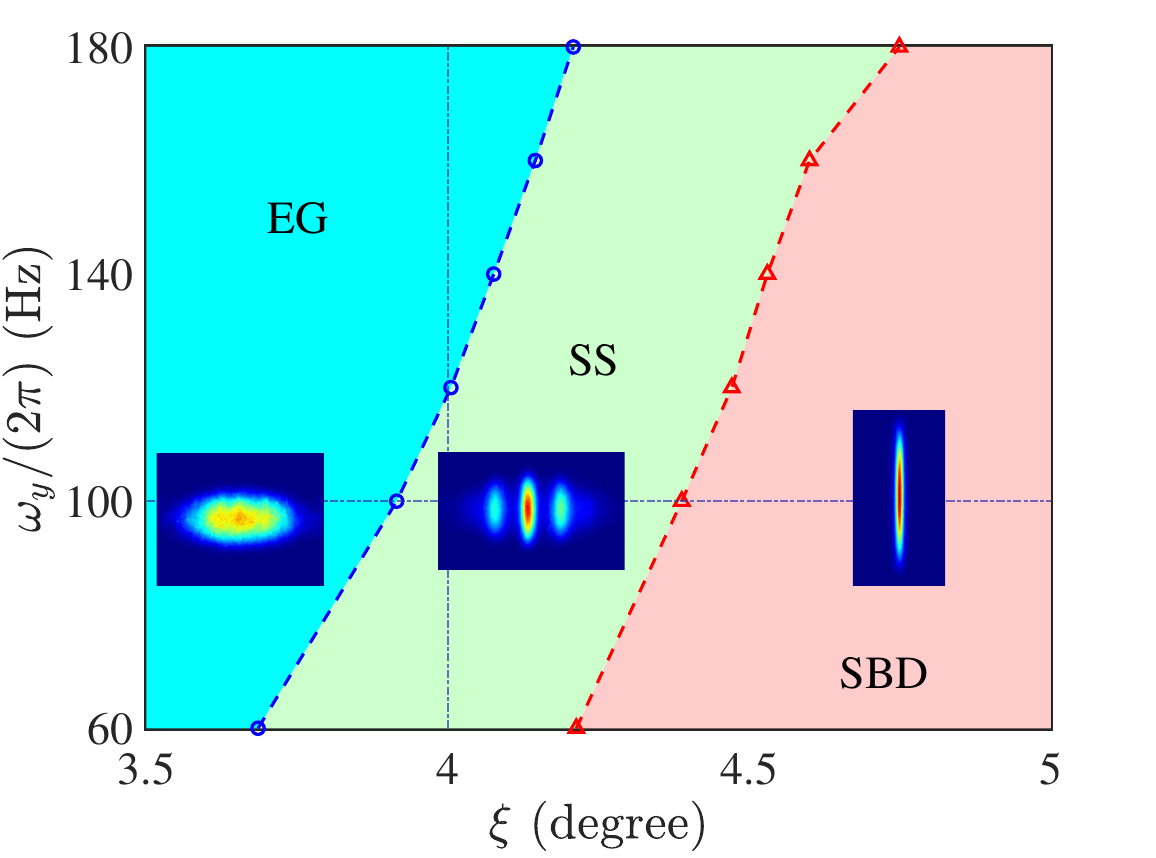}
\caption{Phase diagram in the $\xi$-$\omega_y$ plane. Insets are the representative column densities $\bar{n}(x,y)$ in different phases.}
\label{Phase_diagram}
\end{figure}

\textit{Many-body quantum phases}.--- To further investigate the many-body quantum phases, we consider a trapped gas of $N$ NaCs molecules at temperature $T$. In second-quantized form, the total Hamiltonian of the system reads
\begin{align}
H&=\int d{\bm{r}}\left[ \frac{\hbar^2}{2M}\nabla \hat{\psi}^{\dagger }({
\bm{r}})\nabla \hat{\psi}({\bm{r}})+U({\bm{r}})\hat{\psi}
^{\dagger }({\bm{r}})\hat{\psi}({\bm{r}})\right]\nonumber\\
&\quad+\frac{1}{2}\int d{\bm{r}}d{\bm{r}}^{\prime }V_{\rm eff}(
\bm{r}-\bm{r}^{\prime })\hat{\psi}^{\dagger }({\bm{r}})\hat{\psi}%
^{\dagger }({\bm{r}}^{\prime })\hat{\psi}({\bm{r}}^{\prime })\hat{%
\psi}({\bm{r}}),
\end{align}%
where $M$ is the molecular mass, $\hat{\psi}({\bm{r}})$ is the field operator, and $U({\bm{r}})=M\left(\omega_{x}^{2}x^{2}+\omega_y^2y^{2}+\omega _{z}^{2}z^{2}\right)/2$ is the external trap with $\omega_{i=x,y,z}$ being the trap frequencies. To be specific, we fix the temperature at $T=4\,\mathrm{nK}$ and the trap frequencies at $\omega _{x}=2\pi \times 20\,\mathrm{Hz}$ and $\omega _{z}=2\pi \times 80\,\mathrm{Hz}$. This reduces the system's free parameters to $\xi$ and $\omega_y$. We explore the finite-temperature many-body phases employing the PIMC based on WA~\cite{prokof1998exact, Boninsegni2006a, Boninsegni2006b}, which provides an unbiased and numerically exact framework for investigating the thermodynamic properties of the system under realistic experimental conditions, circumventing the need for mean-field approximations or uncontrolled expansions.

\begin{figure*}[tbp]
\includegraphics[trim=30 60 0 0, clip,width=1.0\linewidth]{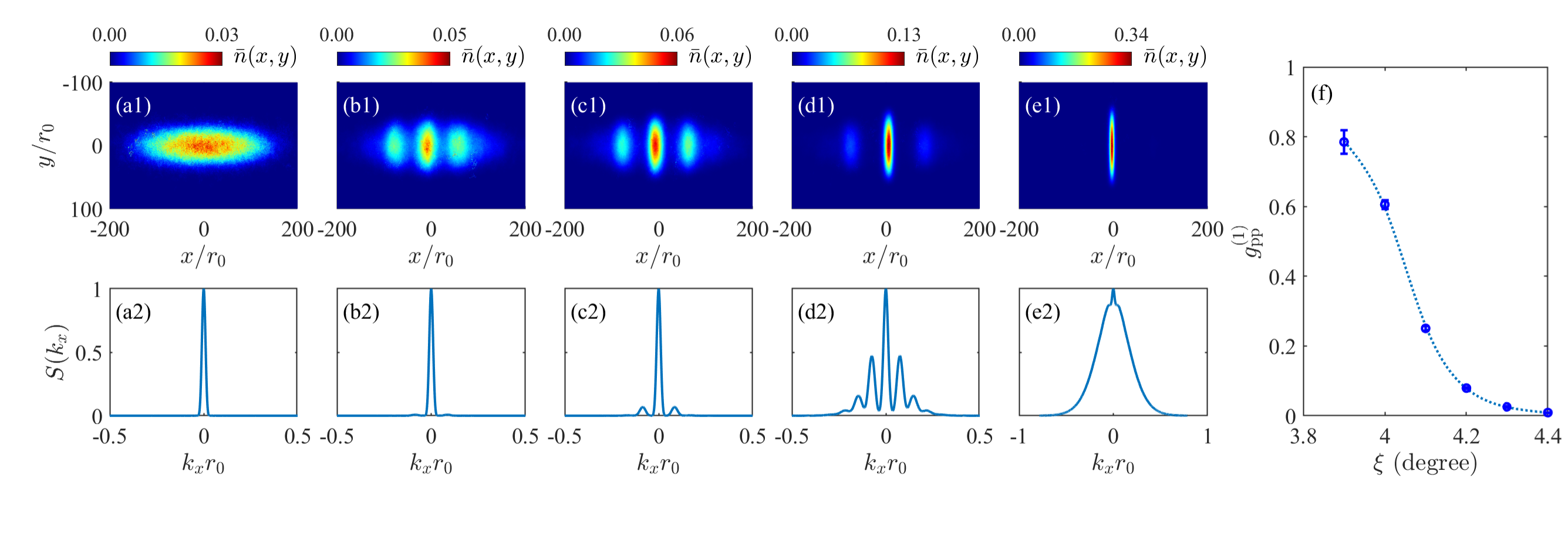}
\caption{Column densities (first row) and structure factors (second row) for $\xi=0^\circ$ (a), $3.9^\circ$ (b), $4.0^\circ$ (c), $4.2^\circ$ (d), and $4.6^\circ$ (e). (f) shows the peak-to-peak coherence versus $\xi$ in the SS phase. Here, column densities are plotted using $10^6 \sim 10^7$ snapshots in the Markov chain generated from a single random seed,
while structure factors and the peak-to-peak coherence are averaged over 40-64 random seeds.}
\label{denSk_xi}
\end{figure*}

\begin{figure*}[tbp]
\includegraphics[trim=30 60 0 0, clip,width=1.0\linewidth]{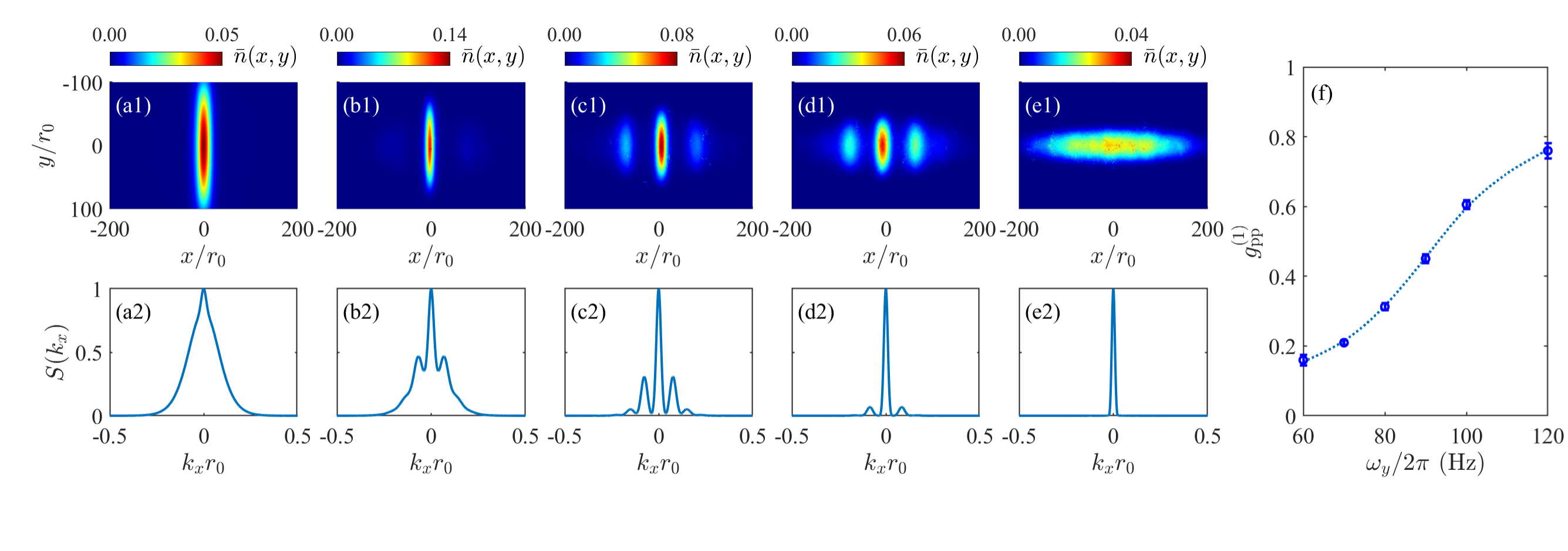}
\caption{Column densities (first row) and structure factors (second row) for $\omega_y/(2\pi)=20$ (a), $60$ (b), $80$ (c), $100$ (d), and $180\,{\rm Hz}$ (e). (f) shows the peak-to-peak coherence versus $\omega_y$ in the SS phase.}
\label{denSk_oy}
\end{figure*}

Figure~\ref{Phase_diagram} summarizes the phase diagram on the $\xi$-$\omega_y$ parameter plane where three distinct phases exist. The insets are representative column densities $\bar{n}(x,y)=\int dzn({\boldsymbol r})$ for different phases. In the EG phase, the cloud is featured by a single density peak that eventually expands and disappears if $\omega_x=0$. The typical condensate fraction $f_c$ is over $0.7$, here $f_c\equiv N_0/N$ with $N_0$ being the largest eigenvalue of the first-order correlation function $G^{(1)}(\bm{r},\bm{r}^{\prime}) \equiv\langle\hat{\psi}^{\dag}(\bm{r})\hat{\psi}(\bm{r}^\prime)\rangle$. The transition from EG to SS phase is determined by the criterion $S(k_{\rm{sub}},0,0)>1\%$, where $S({\boldsymbol k})=\left|\int d{\boldsymbol r}n({\boldsymbol r})e^{-i{\boldsymbol k}\cdot{\boldsymbol r}}\right/N|^2$ is the structure factor, and $k_{\rm{sub}}\neq0$ is the momentum at the sub-dominated peak along the $x$-direction [cf. second rows of Figs.~\ref{denSk_xi} and~\ref{denSk_oy}]. In the SS phase, the gas density along the $x$ axis is modulated such that the whole gas possesses multiple density peaks. Meanwhile, the system remains fully coherent, which gives rise to superfluidity. We remark that since in the presence of density modulation it is challenging to diagonalize $G^{(1)}({\boldsymbol{r}},{\boldsymbol{r}}^{\prime})$ with sufficient precision, we measure the peak-to-peak coherence using the reduced normalized correlation function:
\begin{equation}
g_{\rm pp}^{(1)}\equiv \frac{{\bar G}^{(1)}(x_p,x_p')}{\sqrt{{\bar G}^{(1)}(x_p,x_p) {\bar G}^{(1)}(x_p',x_p')}}
\end{equation}
where ${\bar G}^{(1)}(x,x')=\int dy\,dz\,dy^{\prime }\,dz^{\prime }\,
G^{(1)}(\boldsymbol{r}, \boldsymbol{r}^{\prime })$, $x_p$ and $x_p'$ are the $x$ coordinates of the dominant central peak and its neighboring peak, respectively. In Fig.~\ref{Phase_diagram}, we adopt the criterion $g_{\rm pp}^{(1)}\leq 1\%$ for the transition from SS to the SBD phase. Finally, in the SBD phase, the peak-to-peak coherence vanishes such that different density peaks become fully separated. Moreover, the dominant central droplet is self-bound in the absence of the $x$-direction trap. We note that away from the SS-SBD boundary, the side peaks may disappear completely such that the whole system becomes a single droplet.

\textit{Transitions induced by ellipticity}.---To reveal more details about the quantum phases, we explore the transitions induced by varying ellipticity with the $y$-direction trap frequency being fixed at $\omega_y=2\pi \times 100\,\mathrm{Hz}$. In Fig.~\ref{denSk_xi}, we plot the column density $\bar n(x,y)$ (first row) and the structure factor $S(k_x,0,0)$ (second row) for the representative values of $\xi$ [from (a) to (e)].

At $\xi = 0^{\circ}$, the system forms a single gas cloud elongated along the $x$ direction, consistent with the trap geometry. Consequently, $S(k_x,0,0)$ is a single narrow peak in the momentum space. As $\xi$ increases, the attraction along the $y$ direction becomes stronger, which, analogous to conventional dipolar gases, can lead to a roton instability and trigger density modulation. To see this, we model the effective interaction using the pseudopotential
\begin{align}
V_{\rm pp}({\boldsymbol r})=g_s\delta({\boldsymbol r})+\frac{C_{3,2}}{r^3}\sin ^{2}\theta \cos 2\varphi,
\end{align}
where $g_s=4\pi\hbar^2a_s/M$ with $a_s$ being the $s$-wave scattering length computed using the full effective potential Eq.~\eqref{effpot}. Then, for a homogeneous condensate of the density $n$, the dispersion relation for the Bogoliubov excitation is
\begin{align}
\epsilon({\boldsymbol k})=\sqrt{\epsilon_k^0\left[\epsilon_k^0+2g_sn\left(1-\varepsilon_{3,2}\sin ^{2}\theta_k \cos 2\varphi_k\right)\right]},\label{disper}
\end{align}
where $\epsilon_k^0=\hbar^2k^2/(2M)$, $\varepsilon_{3,2}=MC_{3,2}/(3\hbar^2 a_s)$, and $\theta_k$ and $\varphi_k$ are the polar and azimuthal angles of ${\boldsymbol k}$, respectively. Apparently, for our system with $\varepsilon_{3,2}>0$, the instability can be most easily induced along the direction $(\theta_k,\varphi_k)=(\pi/2,0)$, i.e., the $x$ axis. Although this instability may seem independent of $k$ from Eq.~\eqref{disper}, it actually sets in at a finite $k_x$ for trapped gases. Indeed, as shown in Fig.~\ref{denSk_xi}(b1) and (b2), our numerical simulations indicate that the density becomes modulated along the $x$ axis at $\xi\approx3.9^\circ$. Here, because different density peaks are not well-separated, one can barely see any additional peaks on $S(k_x,0,0)$ other than the one at $k_x=0$. We note that the peak density ($\sim1.5\times 10^{12} \,{\rm cm}^{-3}$) of the gas in Fig.~\ref{denSk_xi}(b1) is at the same order of magnitude as that of NaCs condensate realized in experiment~\cite{Will2024} and is much lower than that of the SS state in Dy atomic gas ($5\times 10^{14}\,{\rm cm}^{-3}$ in~\cite{metasupersolid2019}).

Independent of phase coherence, the density modulation can be alternatively understood from the long-range repulsive interaction along the $x$ direction. Roughly speaking, when the center of the trap is occupied, it is energetically favorable for its immediate neighbors along the $x$ axis to be less occupied, while more distant regions become more populated. Moreover, the attractive interaction along the $y$ axis also facilitates the formation of the density modulation, since all density peaks now have a shape more stretched along the $y$ axis, which further lowers the interaction energy. Finally, the weak $x$-direction confinement further assists the formation of the crystalline order by permitting gas expansion along the axis with minimal potential energy cost. Therefore, the appearance of the SS phase at a relatively low density is due to the interplay between the anisotropies of the interaction and the confinement.

Now, for $\xi=4.0^{\circ}$ [Fig.~\ref{denSk_xi}(c1) and (c2)], well-separated density peaks form. Consequently, the structure factor is characterized by two side peaks at $k_{x}r_{0}=\pm 0.0847(6)$, a signature of the density modulation. When $\xi$ is further increased to $4.2^\circ$, as shown in Fig.~\ref{denSk_xi}(d1), the central peak becomes more stretched along the $y$ axis and accumulates more molecules with a dramatically increased peak density. Such configuration yields a much lower interaction energy. In the momentum space, as shown in Fig.~\ref{denSk_xi}(d2), more side peaks emerge in $S(k_x,0,0)$. Particularly, the subdominant peaks shift to $k_x r_0 = \pm 0.0758(7)$. To confirm the superfluidity in the presence of density modulation, we plot, in Fig.~\ref{denSk_xi}(f), the $\xi$ dependence of the peak-to-peak coherence $g_{\rm pp}^{(1)}$. As can be seen, the coherence monotonically decreases with $\xi$ and becomes negligibly small at $\xi=4.4^\circ$, marking the boundary of the SS phase. We point out that the superfluidity of the gas can also be confirmed by examining the superfluid fraction $f_{\rm sf}^{(i)}$ evaluated in PIMC-WA through the system's response to an imposed rotation around the $i$th axis. The results are presented in SM~\cite{SM}. Because the system is highly anisotropic, $f_{\rm sf}^{(i)}$ for different axes are distinct. In addition, since none of $f_{\rm sf}^{(i)}$ exactly measure the superfluidity along the $x$ axis, they remain finite even when the coherence vanishes.

Finally, in the droplet phase with $\xi=4.6^\circ$, as shown in Fig.~\ref{denSk_xi}(e1), all side peaks merge into the central one such that it becomes a filament with a peak density ten times larger than that of the $\xi = 0^{\circ}$ case. Correspondingly, $S(k_x,0,0)$ exhibits a broadened peak [Fig.~\ref{denSk_xi}(e2)], indicating the loss of the crystalline order.

\textit{Transitions induced by confinement}.---We now turn to explore the phase transitions induced by varying $\omega_y$. For this purpose, we fix the ellipticity at $\xi=4^\circ$ and allow $\omega_y/(2\pi)$ to vary between $20$ to $180\,{\rm Hz}$. Figure~\ref{denSk_oy} plots the typical column densities $\bar n(x,y)$ (first row) and structure factors $S(k_x,0,0)$ (second row) for different values of $\omega_y$. 

For $\omega_y/(2\pi)=20\,{\rm Hz}$, the trapping potential is isotropic in the $xy$ plane and is more tightly confined along the $z$ axis. However, due to the anistropy of the interaction, the gas, as shown in Fig.~\ref{denSk_oy}(a1), is dramatically stretched along the $y$ axis. Accordingly, $S(k_x,0,0)$ exhibits a single broad peak in momentum space [see Fig.~\ref{denSk_oy}(a2)]. Moreover, since the system is self-bound, it belongs to the SBD phase. The condensate fraction is $f_c =0.63(1)$. 

Following the increase of $\omega_y$, the confining potential squeezes the gas in the $y$ direction. To keep the interaction energy low, the gas roughly maintains its shape and shrinks as a whole. In consequence, the peak density of the gas increases significantly. As the inter-molecular spacing becomes comparable to the size of the shielding potential, further increase of the peak density becomes infeasible. Instead, the gas starts to spread along the $x$ axis which has the weakest confinement. When the gas expands sufficiently along the repulsive $x$ direction, side density peaks develop to lower the interaction energy, which leads to the crystalline order.

Figure~\ref{denSk_oy}(b), (c), and (d) show the column density and the structure factor in the SS phase for $\omega_y/(2\pi)=60$, $80$, and $100\,{\rm Hz}$, respectively. The crystalline order is clearly manifested in the structure factor, where the subdominant peaks shift from $k_x r_0 = \pm 0.0652(7)$ to $\pm 0.0764(3)$ and then to $\pm 0.0847(6)$ as $\omega_y$ increases. In addition, the central peak is continuously squeezed, accompanied by the growth of the side peaks. Consequently, the crystalline order becomes well formed in Fig.~\ref{denSk_oy}(c). When more molecules are squeezed into the side peaks [see Fig.~\ref{denSk_oy}(d)], the boundary between the central and side peaks is blurred and the crystalline order is suppressed. In other words, the supersolid starts to melt. In Fig.~\ref{denSk_oy}(f), we plot $g_{\rm pp}^{(1)}$ as a function of $\omega_y$. As can be seen, the peak-to-peak coherence monotonically increases with $\omega_y$, which is understandable since the peak density of the gas decreases. Finally, as $\omega_y/(2\pi)$ increases to $180\,{\rm Hz}$, the system enters the EG phase for which, as shown in Fig.~\ref{denSk_oy}(e), the side peaks disappear completely.

\textit{Conclusion}.---For ultracold gases of polar molecules shielded by dual microwaves, we derive an effective potential that is anisotropic on the $xy$ plane by introducing ellipticity. We demonstrate that by tuning the microwave ellipticity, the SS phase can be realized in ultracold gases of NaCs molecules. We map out the phase diagram on the $\xi$-$\omega_y$ plane which shows that the SS phase lies in the parameter regime accessible to current experiments. We have analyzed the underlying physics for the emergence of the density modulation, which shows that through the interplay between anisotropies of interaction and confinement the SS phase can be realized in a gas with relatively low density. 

\begin{acknowledgments}
This work was supported by National Key Research and Development Program of China (Grant No. 2021YFA0718304), by the NSFC (Grants No. 12135018, No. 12047503, No. 12474245, and No. 12274331), and by CAS Project for Young Scientists in Basic Research (Grant No. YSBR-057).
\end{acknowledgments}

\bibliography{ref_dMolecule.bib}

\clearpage

\widetext

\begin{center}
\textbf{\large Supplemental Materials}
\end{center}

This Supplemental Material is structured as follows. In Section~\ref{appmodel}, we present a model for polar molecules subjected to both an elliptically polarized and a linearly polarized microwave field, leading to a time-dependent two-molecule Hamiltonian. In Section~\ref{Floqeff}, we employ Floquet theory to derive the effective interaction potential for two molecules in the highest dressed state, which shows excellent agreement with the adiabatic potential. Section~\ref{appflosc} validates this effective potential through multi-channel scattering calculations. In Section~\ref{SF}, we demonstrate the anisotropic superfluid fractions by changing the ellipticity and trap frequency. 

\setcounter{equation}{0} \setcounter{figure}{0} \setcounter{table}{0} %
\setcounter{page}{1} \setcounter{section}{0} \makeatletter
\renewcommand{\theequation}{S\arabic{equation}} \renewcommand{\thefigure}{S%
\arabic{figure}} \renewcommand{\bibnumfmt}[1]{[S#1]} \renewcommand{%
\citenumfont}[1]{S#1} \renewcommand{\thesection}{S\arabic{section}}%
\setcounter{secnumdepth}{3}

\renewcommand{\thefigure}{SM\arabic{figure}} \renewcommand{\thesection}{SM
\arabic{section}} \renewcommand{\theequation}{SM\arabic{equation}}

\section{Single-molecule eigenstates and two-body interactions}
\label{appmodel}

\begin{figure}[tbp]
\includegraphics[width=0.5\linewidth]{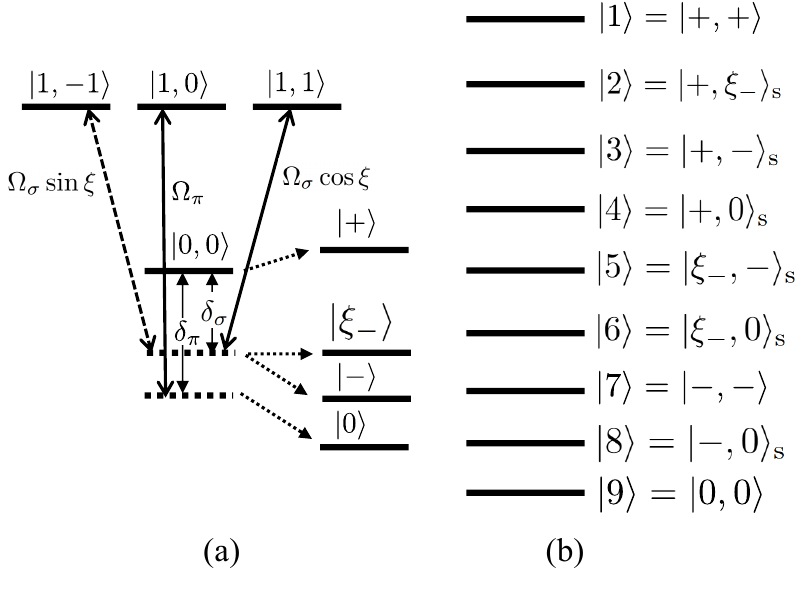}
\caption{Schematics of the dressed-state energy levels. (a) Single-molecule
states. (b) Two-molecule states in the symmetric subspace.}
\label{levels}
\end{figure}

We consider an ultracold gas of bialkali polar molecules in the $^{1}\Sigma
(v=0)$ state, which can be treated as rigid rotors. At ultracold
temperatures, it is sufficient to focus on the lowest rotational state $%
|J,M_{J}\rangle =|0,0\rangle $ and the first excited manifold $%
|J=1,M_{J}=0,\pm 1\rangle $, separated by an energy $\hbar \omega _{e}$.
Each molecule possesses a permanent electric dipole moment $d\hat{%
\boldsymbol{d}}$, oriented along the internuclear axis represented by the
unit vector $\hat{\boldsymbol{d}}$. To maintain the shielding effect, we apply an elliptically polarized microwave field with elliptic angle $\xi$, along with a $\pi $-polarized microwave. As illustrated in Fig.~\ref{levels}, these fields couple the transitions $|0,0\rangle \leftrightarrow
|1,\pm 1\rangle $ and $|0,0\rangle \leftrightarrow |1,0\rangle $, with
Rabi-frequencies $\Omega _{\sigma }$ and $\Omega _{\pi }$, respectively. The
corresponding microwave frequencies $\omega _{\sigma }$ and $\omega _{\pi }$
are blue detuned from the rotational transition frequency $\omega _{e}$,
with detunings $\delta _{\sigma ,\pi }=\omega _{e}-\omega _{\sigma ,\pi }$.

In the frame co-rotating with microwave fields, the internal-state
Hamiltonian for a single molecule is time-independent and takes the form
\begin{eqnarray}
\hat{h}_{\mathrm{in}} &=&\delta _{\sigma }\Big(|\xi _{+}\rangle \langle \xi
_{+}|+|\xi _{-}\rangle \langle \xi _{-}|\Big)+\delta _{\pi }|1,0\rangle
\langle 1,0|  \notag \\
&&+\left( \frac{\Omega _{\sigma }}{2}|\xi _{+}\rangle \langle 0,0|+\frac{
\Omega _{\pi }}{2}|1,0\rangle \langle 0,0|+\mathrm{H.c.}\right),
\end{eqnarray}
where $\left\vert \xi _{+}\right\rangle =\cos \xi
|1,1\rangle +\sin \xi |1,-1\rangle $, and $\left\vert \xi _{-}\right\rangle
=\cos \xi |1,-1\rangle -\sin \xi |1,1\rangle $ are determined by the
elliptic angle of the microwave. In the basis $\left\{ \left\vert
0,0\right\rangle ,\left\vert \xi _{+}\right\rangle ,\left\vert
1,0\right\rangle ,\left\vert \xi _{-}\right\rangle \right\} $, the
single-molecule Hamiltonian reads
\begin{equation}
\hat{h}_{\mathrm{in}}=
\begin{pmatrix}
0 & \frac{\Omega _{\sigma }}{2} & \frac{\Omega _{\pi }}{2} & 0 \\ 
\frac{\Omega _{\sigma }}{2} & \delta _{\sigma } & 0 & 0 \\ 
\frac{\Omega _{\pi }}{2} & 0 & \delta _{\pi } & 0 \\ 
0 & 0 & 0 & \delta _{\sigma }
\end{pmatrix}
,
\end{equation}

The single-molecule Hamiltonian $\hat{h}_{\mathrm{in}}$ can be analytically
diagonalized by an unitary transformation
\begin{equation}
U_{1}=
\begin{pmatrix}
\cos \alpha & 0 & -\sin \alpha \cos \gamma & \sin \alpha \sin \gamma \\ 
\sin \alpha \cos \beta & 0 & \cos \alpha \cos \beta \cos \gamma -\sin \beta
\sin \gamma & -\cos \alpha \cos \beta \sin \gamma -\sin \beta \cos \gamma \\ 
\sin \alpha \sin \beta & 0 & \cos \alpha \sin \beta \cos \gamma +\cos \beta
\sin \gamma & -\cos \alpha \sin \beta \sin \gamma +\cos \beta \cos \gamma \\ 
0 & 1 & 0 & 0
\end{pmatrix}
,
\end{equation}
which is conveniently parameterized by three Euler angles $\alpha $, $\beta $
, and $\gamma $. The columns of $U_{1}$ are the eigenvectors of $\hat{h}_{
\mathrm{in}}$ which, from left to right, are denoted as $\left\vert
+\right\rangle $, $\left\vert \xi _{-}\right\rangle $, $\left\vert
-\right\rangle $, and $\left\vert 0\right\rangle $. The corresponding
eigenenergies are denoted as $E_{+}$, $\delta _{\sigma }$, $E_{-}$, and $
E_{0}$, respectively. We note that in the limit $\Omega _{\pi }\rightarrow 0$
the eigenstates and eigenenergies can be expressed explicitly as $|+\rangle
\rightarrow \cos \alpha \left\vert 0,0\right\rangle +\sin \alpha \left\vert
\xi _{+}\right\rangle $, $|-\rangle \rightarrow \sin \alpha \left\vert
0,0\right\rangle -\cos \alpha \left\vert \xi _{+}\right\rangle $, $|0\rangle
\rightarrow |1,0\rangle $, $E_{\pm }\rightarrow (\delta _{\sigma }\pm \Omega
_{\mathrm{eff}})/2$, and $E_{0}\rightarrow \delta _{\pi }$, where $\Omega _{
\mathrm{eff}}=\sqrt{\delta _{\sigma }^{2}+\Omega _{\sigma }^{2}}$ and the
Euler angles are known analytically, i.e., $(\alpha ,\beta ,\gamma )=\left(
\arccos [(1-\delta _{\sigma }/\Omega _{\mathrm{eff}})/2]^{1/2},0,0\right) $.
The energy level structure in the single-molecule subspace is shown in Fig.~
\ref{levels}a.

Now, for two molecules with dipole moments $d\hat{\boldsymbol{d}}_{1}$ and $d
\hat{\boldsymbol{d}}_{2}$, the inter-molecular DDI is
\begin{align}
V_{\mathrm{dd}}({\boldsymbol{r}})& =\frac{d^{2}}{4\pi \epsilon _{0}r^{3}}
\left[ \hat{\boldsymbol{d}}_{1}\cdot \hat{\boldsymbol{d}}_{2}-3(\hat{
\boldsymbol{d}}_{1}\cdot \hat{\boldsymbol{r}})(\hat{\boldsymbol{d}}_{2}\cdot 
\hat{\boldsymbol{r}})\right]  \notag \\
& =-\frac{\eta }{r^{3}}\sum_{m=-2}^{2}Y_{2m}^{\ast }(\hat{\boldsymbol{r}}
)\Sigma _{2,m},
\end{align}
where $\eta =\sqrt{8\pi /15}\,d^{2}/\epsilon _{0}$ with $\epsilon _{0}$
being the electric permittivity of vacuum, $r=|{{\boldsymbol{r}}}|$, $Y_{2m}(
\hat{\boldsymbol{r}})$ are spherical harmonics, and $\Sigma _{2,m}$ are
components of the rank-2 spherical tensor defined as $\Sigma _{2,0}=(\hat{d}
_{1}^{+}\hat{d}_{2}^{-}+\hat{d}_{1}^{-}\hat{d}_{2}^{+}+2\hat{d}_{1}^{0}\hat{d
}_{2}^{0})/\sqrt{6}$, $\Sigma _{2,\pm 1}=(\hat{d}_{1}^{\pm }\hat{d}_{2}^{0}+
\hat{d}_{1}^{0}\hat{d}_{2}^{\pm })/\sqrt{2}$, and $\Sigma _{2,\pm 2}=\hat{d}
_{1}^{\pm }\hat{d}_{2}^{\pm }$ with $\hat{d}_{j}^{\pm }=Y_{1,\pm 1}(\hat{
\boldsymbol{d}}_{j})$ and $\hat{d}_{j}^{0}=Y_{1,0}(\hat{\boldsymbol{d}}_{j})$
. In the rotating frame, $\hat{d}_{j}^{\pm }$ and $\hat{d}_{j}^{0}$ become
time-dependent. and in the basis $|J,M_{J}\rangle $ these operators can be
written out explicitly as
\begin{align}
\hat{d}_{j}^{0}& =\frac{1}{\sqrt{4\pi }}\left( |0,0\rangle \langle
1,0|e^{-i\omega _{\pi }t}+\mathrm{h.c.}\right) ,  \notag \\
\hat{d}_{j}^{+}& =\frac{1}{\sqrt{4\pi }}\left( |1,1\rangle \langle
0,0|e^{i\omega _{\sigma }t}-|0,0\rangle \langle 1,-1|e^{-i\omega _{\sigma
}t}\right) ,  \notag \\
\hat{d}_{j}^{-}& =-\left( \hat{d}_{j}^{+}\right) ^{\dag }.
\end{align}
After substituting the above spherical components of the vector into $\Sigma
_{2m}$, we find 
\begin{align}
\Sigma _{2,0} &=\frac{1}{4\pi \sqrt{6}}\Big(2\left\vert 1,0\right\rangle
\left\langle 0,0\right\vert \otimes \left\vert 0,0\right\rangle \left\langle
1,0\right\vert -\left\vert 1,1\right\rangle \left\langle 0,0\right\vert
\otimes \left\vert 0,0\right\rangle \left\langle 1,1\right\vert -\left\vert
0,0\right\rangle \left\langle 1,-1\right\vert \otimes \left\vert
1,-1\right\rangle \left\langle 0,0\right\vert +\mathrm{h.c.}\Big),  \notag \\
\Sigma _{2,1}&=\frac{1}{4\pi \sqrt{2}}\Big[\big(\left\vert 1,1\right\rangle
\left\langle 0,0\right\vert \otimes \left\vert 0,0\right\rangle \left\langle
1,0\right\vert +\left\vert 0,0\right\rangle \left\langle 1,0\right\vert
\otimes \left\vert 1,1\right\rangle \left\langle 0,0\right\vert \big)
e^{i\omega t}  \notag \\
&\qquad\qquad\;-\big(\left\vert 0,0\right\rangle \left\langle
1,-1\right\vert \otimes \left\vert 1,0\right\rangle \left\langle
0,0\right\vert +\left\vert 1,0\right\rangle \left\langle 0,0\right\vert
\otimes \left\vert 0,0\right\rangle \left\langle 1,-1\right\vert \big)
e^{-i\omega t}\Big],  \notag \\
\Sigma _{2,2} &=-\frac{1}{4\pi }\big(\left\vert 1,1\right\rangle
\left\langle 0,0\right\vert \otimes \left\vert 0,0\right\rangle \left\langle
1,-1\right\vert +\left\vert 0,0\right\rangle \left\langle 1,-1\right\vert
\otimes \left\vert 1,1\right\rangle \left\langle 0,0\right\vert \big), 
\notag
\end{align}
where, according to the rotating-wave approximation, we have neglected
time-dependent terms with higher frequencies $\omega _{+}$ and $\omega _{\pi
}$ (of the order of $\mathrm{GHz}$) and retained those with the lower
frequency $\omega =\omega _{\sigma }-\omega _{\pi }$ (of the order of MHz typically).

To proceed further, the Hamiltonian for the relative motion of two molecules
is
\begin{equation*}
\hat{H}_{2}=-\frac{\hbar ^{2}\nabla ^{2}}{M}+\sum_{j=1,2}\hat{h}_{\mathrm{in}
}(j)+V_{\mathrm{dd}}({\boldsymbol{r}},t),
\end{equation*}
where $M$ is the mass of the molecule and $\hat{h}_{\mathrm{in}}(j)$ denotes
the internal-state Hamiltonian of the $j$th molecule. And we have explicitly
expressed $V_{\mathrm{dd}}$ as a function of $t$ in the rotating frame.
Since $\hat{H}_{2}$ possesses a parity symmetry, the symmetric and
antisymmetric two-particle internal states are decoupled in the Hamiltonian $
\hat{H}_{2}$. We shall only focus on the ten-dimensional symmetric subspace
in which the microwave shielded two-molecule state $|1\rangle \equiv
|+\rangle \otimes |+\rangle $ lies. Further simplification can be made by
noting that $|1\rangle $ only couples to the following eight two-molecule
states: $|2\rangle \equiv |+,\xi _{-}\rangle _{\mathrm{s}}$, $|3\rangle
\equiv |+,-\rangle _{\mathrm{s}}$, $|4\rangle \equiv |+,0\rangle _{\mathrm{s}
}$, $|5\rangle \equiv |\xi _{-},-\rangle _{\mathrm{s}}$, $|6\rangle \equiv
|\xi _{-},0\rangle _{\mathrm{s}}$, $|7\rangle \equiv |-\rangle \otimes
|-\rangle $, $|8\rangle \equiv|-,0\rangle _{\mathrm{s}}$, and $|9\rangle \equiv
|0\rangle \otimes |0\rangle $, where $|i,j\rangle _{\mathrm{s}}=(|i\rangle
\otimes |j\rangle +|j\rangle \otimes |i\rangle )/\sqrt{2}$ represents the
symmetrized two-molecule state. The corresponding energies of these
two-particle states are denoted as $E_{\nu =1\sim 9}^{(\infty )}$. As a
result, these nine two-molecule states form a 9-dimensional (9D) symmetric
subspace, $\mathcal{S}_{9}\equiv \mathrm{span}\{|\nu \rangle \}_{\nu =1}^{9}$
. Now, since we focus on system with all molecules being prepared in the
microwave shielded $|+\rangle $ state, we may project the interaction $V_{
\mathrm{dd}}({\boldsymbol{r}})$ onto the two-molecule subspace $\mathcal{S}
_{9}$. The energy level structure in the two-molecule subspace in the limit $
r \to \infty$ is shown in Fig.~\ref{levels}b.

In an attempt to eliminate the time dependence of the $\hat{H}_{2}$, we
introduce, in $\mathcal{S}_{9}$, an unitary transformation defined by the
diagonal matrix
\begin{equation}
U_{2}(t)=\mathrm{diag}\left( 1,1,1,e^{-i\omega t},1,e^{-i\omega
t},1,e^{-i\omega t},e^{-2i\omega t}\right) .
\end{equation}
A straightforward calculation shows that the Hamiltonian $\hat{H}_{2}$ is
transformed into 
\begin{eqnarray}
\hat{\mathcal{H}}(t) &=&U_{2}^{\dag }(t)\hat{H}_{2}U_{2}(t)-iU_{2}^{\dagger
}(t)\partial _{t}U_{2}(t)  \notag \\
&=&-\frac{\hbar ^{2}\nabla ^{2}}{M}+\mathcal{E}^{(\infty )}+\mathcal{V}({{
\boldsymbol{r}}},t),
\end{eqnarray}
where $\mathcal{E}^{(\infty )}$ is a diagonal matrix with elements being the
energies of the asymptotical state $|\nu \rangle $ with respect to that of
the $|\nu =1\rangle $ state, i.e., $\mathcal{E}_{\nu \nu ^{\prime
}}^{(\infty )}=(E_{\nu }^{(\infty )}-2E_{+})\delta _{\nu \nu ^{\prime }}$.
Moreover, the two-body interaction in the basis $\{|\nu \rangle \}$ is
\begin{equation*}
\mathcal{V}({\boldsymbol{r}},t)=U_{2}^{\dag }(t)V_{\mathrm{dd}}({\boldsymbol{
r}})U_{2}(t)=\sum_{s=-3}^{3}\mathcal{V}_{s}({\boldsymbol{r}})e^{is\omega t}
\end{equation*}
which is decomposed into components according to the time dependence $
e^{is\omega t}$. Particularly, the components satisfy $\mathcal{V}_{-s}({
\boldsymbol{r}})=\mathcal{V}_{s}^{\dagger }({\boldsymbol{r}})$ and

\begin{subequations}
\label{2bodyint}
\begin{align}
\mathcal{V}_{0}({\boldsymbol{r}})& =-\frac{\eta }{r^{3}}\left[ Y_{20}(\hat{
\boldsymbol{r}})\Sigma _{2,0}^{(0)}+Y_{21}^{\ast }(\hat{\boldsymbol{r}}
)\Sigma _{2,1}^{(0)}+Y_{21}(\hat{\boldsymbol{r}})\Sigma _{2,1}^{(0)\dagger
}\right.\left. +Y_{22}^{\ast }(\hat{\boldsymbol{r}})\Sigma
_{2,2}^{(0)}+Y_{22}(\hat{\boldsymbol{r}})\Sigma _{2,2}^{(0)\dagger }\right] \\
\mathcal{V}_{1}({\boldsymbol{r}})& =-\frac{\eta }{r^{3}}\left[ Y_{20}(\hat{
\boldsymbol{r}})\Sigma _{2,0}^{(1)}+Y_{21}(\hat{\boldsymbol{r}})\Sigma
_{2,1}^{(-1)\dag }+Y_{21}^{\ast }(\hat{\boldsymbol{r}})\Sigma
_{2,1}^{(1)}\right.\left. +Y_{22}(\hat{\boldsymbol{r}})\Sigma
_{2,2}^{(-1)\dag }+Y_{22}^{\ast }(\hat{\boldsymbol{r}})\Sigma _{2,2}^{(1)}
\right] , \\
\mathcal{V}_{2}({\boldsymbol{r}})& =-\frac{\eta }{r^{3}}\left[ Y_{20}(\hat{
\boldsymbol{r}})\Sigma _{2,0}^{(2)}+Y_{21}(\hat{\boldsymbol{r}})\Sigma
_{2,1}^{(-2)\dag }+Y_{21}^{\ast }(\hat{\boldsymbol{r}})\Sigma
_{2,1}^{(2)}\right.\left. +Y_{22}^{\ast }(\hat{\boldsymbol{r}})\Sigma
_{2,2}^{(2)}+Y_{22}(\hat{\boldsymbol{r}})\Sigma _{2,2}^{(-2)\dagger }\right]
, \\
\mathcal{V}_{3}({\boldsymbol{r}})& =-\frac{\eta }{r^{3}}\left[ Y_{21}^{\ast
}(\hat{\boldsymbol{r}})\Sigma _{2,1}^{(3)}+Y_{21}(\hat{\boldsymbol{r}}
)\Sigma _{2,1}^{(-3)\dagger }\right] ,
\end{align}
\end{subequations}
where $\Sigma _{2,m}^{(s)}$ are matrices originated from $\Sigma
_{2,m}=\sum_{s}\Sigma _{2,m}^{(s)}e^{is\omega t}$ that is associated with
the time dependence $e^{is\omega t}$, which satisfy $\Sigma
_{2,-m}^{(s)}=(-1)^{m}\Sigma _{2,m}^{(-s)\dag }$.

\section{Floquet theory and effective potentials}
\label{Floqeff}

The time periodicity of the Hamiltonian $\hat{\mathcal{H}}(t)$ suggests that
we may tackle the two-molecule physics using the Floquet theory.
Specifically, the solution of the Schr\"{o}dinger equation, 
\begin{equation}
i\hbar \frac{\partial |\psi (t)\rangle }{\partial t}=\hat{\mathcal{H}}
(t)\left\vert \psi (t)\right\rangle ,  \label{se}
\end{equation}
takes the \textquotedblleft Floquet-Fourier\textquotedblright\ form
\begin{equation}
\left\vert \psi (t)\right\rangle =e^{-i\varepsilon t}\sum_{n=-\infty
}^{\infty }e^{-in\omega t}\left\vert \psi _{n}\right\rangle ,
\end{equation}
where $\varepsilon $ is the quasi-energy of the state and $|\psi _{n}\rangle 
$ is the time-independent harmonic component defined on the 9D Hilbert space 
$\mathcal{S}_{9}$. It follows from Eq.~(\ref{se}) that $|\psi _{n}\rangle $
satisfy the time-independent eigenvalue equation:
\begin{equation}
\sum_{s}\mathcal{H}_{s}\left\vert \psi _{n+s}\right\rangle -n\omega |\psi
_{n}\rangle =\varepsilon \left\vert \psi _{n}\right\rangle ,\quad n=0,\pm
1,\pm 2,\ldots ,  \label{FE}
\end{equation}
where the Floquet Hamiltonian is
\begin{equation}
\mathcal{H}_{s}=\left\{ 
\begin{array}{ll}
-\hbar ^{2}\nabla ^{2}/M+\mathcal{E}^{(\infty )}+\mathcal{V}_{0}({
\boldsymbol{r}}), & \mbox{for }s=0; \\ 
\mathcal{V}_{s}({\boldsymbol{r}}), & \mbox{for }0<|s|\leq 3; \\ 
0, & \mbox{otherwise}.
\end{array}
\right.
\end{equation}
Introducing vector $|{\boldsymbol{\Psi }}\rangle =(\cdots ,|\psi
_{-1}\rangle ,|\psi _{0}\rangle ,|\psi _{1}\rangle ,\cdots )^{T}$ for the
Floquet space wavefuction, Eq.~\eqref{FE} can be rewritten into a more
compact form as 
\begin{equation*}
{\mathbf{H}}|{\boldsymbol{\Psi }}\rangle =\varepsilon |{\boldsymbol{\Psi }}
\rangle ,
\end{equation*}
where, in terms of $9\times 9$ block matrices, the Floquet Hamiltonian ${
\mathbf{H}}$ is a heptadiagonal matrix. Thus one may easily visualize the
structure of the Schr\"{o}dinger Eq.~\eqref{FE} through $\mathbf{H}$.

With Eq.~\eqref{FE}, the advantage of performing the transformation $
U_{2}(t) $ is now understandable. Specifically, in the absence of the $\pi $
-field, $U_{2}(t)$ leads to a time-independent $\hat{\mathcal{H}}$, i.e., $
\mathcal{H}_{s}=0$ for all $s\neq 0$, indicating that different Floquet
sectors are decoupled and the eigenstates can be obtain directly by
diagonalizing $\mathcal{H}_{0}$. Then as the $\pi $-polarized microwave is
gradually turned on to lower the attractive interaction on the $xy$ plane,
the transitions between different Floquet sectors are also switched on. As a result, through transformation $
U_{2}$, we ensure that the transitions between different Floquet sectors in
the presence of the $\pi $ microwave are perturbation. Such structure
accelerates the convergence in numerical calculations.

Since the microwave-shielded $|+\rangle $ state has a sufficiently long
lifetime in experiments, a molecular gas prepared in the $|+\rangle $ state
represents an important platform for studying the many-body physics. We can
derive an time-independent effective potential to describe two-body
scatterings in the $|+\rangle $ state. We note that the internal-state
dynamics is much faster than the motion of molecules, which allows us to
employ the Born-Oppenheimer (BO) approximation. Then, for a given ${
\boldsymbol{r}}$, we diagonalize, in the Floquet space, the potential matrix 
${\mathbf{V}}$ (i.e., the Floquet Hamiltonian $\mathbf{H}$ with kinetic
energy being neglected), which gives rise to
\begin{equation}
{\mathbf{V}}({\boldsymbol{r}})\left\vert V_{n,\nu }^{(\mathrm{ad})}({
\boldsymbol{r}})\right\rangle =V_{n,\nu }^{(\mathrm{ad})}({\boldsymbol{r}}
)\left\vert V_{n,\nu }^{(\mathrm{ad})}({\boldsymbol{r}})\right\rangle ,
\end{equation}
where $|V_{n,\nu }^{(\mathrm{ad})}({\boldsymbol{r}})\rangle $ is the
eigenstate and $V_{n,\nu }^{(\mathrm{ad})}({\boldsymbol{r}})$ is the
eigenenergy. Physically, $|V_{n,\nu }^{(\mathrm{ad})}({\boldsymbol{r}}
)\rangle $ is the state adiabatically connects to the asymptotical state $
|n,\nu \rangle $, i.e., the two-molecule state $|\nu \rangle $ in the $n$-th
Floquet sector. In particular, we focus on the eigenstate adiabatically
connecting to $\left\vert 1\right\rangle =|++\rangle $ with $n=0$ as $
r\rightarrow \infty $. The corresponding eigenenergy $V_{0,1}^{(\mathrm{ad}
)}({\boldsymbol{r}})$ is then the effective potential between two
microwave-shielded molecules.

Alternatively, we may analytically derive an highly accurate effective
potential, $V_{\mathrm{eff}}({\boldsymbol{r}})$, through perturbation
theory. For this purpose, we note that the first-order correction to the
energy of the $\left\vert 1\right\rangle $ state in the sector $n=0$ is
\begin{equation}
\lbrack \mathcal{V}_{0}({\boldsymbol{r}})]_{11}=\langle 1|V_{\mathrm{dd}}({
\boldsymbol{r}})|1\rangle =\frac{1}{r^{3}}[C_{3,0}(3\cos ^{2}\theta
-1)+C_{3,2}\sin ^{2}\theta \cos 2\varphi ],
\end{equation}
where
\begin{eqnarray}
C_{3,0} &=&\sqrt{\frac{15}{2\pi }}\frac{\eta }{48\pi }(3\cos 2\beta -1)\cos
^{2}\alpha \sin ^{2}\alpha ,  \notag \\
C_{3,2} &=&\sqrt{\frac{15}{2\pi }}\frac{\eta }{8\pi }\sin ^{2}\alpha \cos
^{2}\alpha \sin 2\xi \cos ^{2}\beta .
\end{eqnarray}
For fixed $\delta _{\sigma }$, $\Omega _{\sigma }$, $\delta _{\pi }$, and $
\xi =0$, the Euler angle $\beta $ increases with $\Omega _{\pi }$. As $
\Omega _{\pi }$ increases to the threshold value $\Omega _{c}$, $\beta $
reaches $\beta _{c}=\arccos (1/3)/2$, resulting in a complete cancellation
of the effective DDI, i.e., $C_{3,0}=0$. Below the completely cancellation
point, e.g., $\Omega _{\pi }<\Omega _{c}$, the elliptic angle $\xi >0$ leads
to $C_{3,2}>0$, which can be applied to enhance (reduce) the attractive
interaction along the $y$ ($x$)-direction.

Next, the second-order correction can be formally expressed as
\begin{equation}
\left. \sum_{s,\nu }\right. ^{\prime }\frac{\left\vert \left( \mathcal{V}
_{s}({\boldsymbol{r}})\right) _{1\nu }\right\vert ^{2}}{\mathcal{E}
_{11}^{(\infty )}-\left[ \mathcal{E}_{\nu \nu }^{(\infty )}-s\omega \right] }
,  \label{2ndcorr}
\end{equation}
where the primed sum excludes the term with $(s,\nu )=(0,1)$. By carefully
examining the matrix elements of $\mathcal{V}_{s}({\boldsymbol{r}})$, it turns out that, in Eq.~\eqref{2ndcorr}, only a
finite number of terms contribute. After collecting all terms contributing
to the second-order correction, the effective potential can now be
approximated as
\begin{eqnarray}
V_{\mathrm{eff}}({\boldsymbol{r}}) &=&\frac{1}{r^{3}}[C_{3,0}(3\cos
^{2}\theta -1)+C_{3,2}\sin ^{2}\theta \cos 2\varphi ]  \notag \\
&&+\frac{\eta ^{2}}{r^{6}}\sum_{mm^{\prime }}C_{6,mm^{\prime }}Y_{2m}^{\ast
}(\hat{r})Y_{2m^{\prime }}(\hat{r}),
\label{Veff}
\end{eqnarray}
where the symmetric matrix $C_{6}$ in the basis $Y_{2,m=2,...,-2}(\hat{r})$
can be obtained analytically in terms of the matrix elements $(\Sigma
_{2,m}^{(s)})_{1\nu }$, $\mathcal{E}_{\nu \nu }^{(\infty )}$, and $\omega $.
The parity symmetry imposes $C_{6,mm}=C_{6,-m,-m}$ and non-vanishing $C_{6,mm'}$ with $m^{\prime }=m,m\pm
2,m\pm 4$, which determines the structure
\begin{equation}
C_{6}=\left( 
\begin{array}{ccccc}
C_{6,22} & 0 & C_{6,20} & 0 & C_{6,2-2} \\ 
0 & C_{6,11} & 0 & C_{6,1-1} & 0 \\ 
C_{6,20} & 0 & C_{6,00} & 0 & C_{6,20} \\ 
0 & C_{6,1-1} & 0 & C_{6,11} & 0 \\ 
C_{6,2-2} & 0 & C_{6,20} & 0 & C_{6,22}
\end{array}
\right) .
\end{equation}

In Fig.~\ref{effpot}(a), we show the interaction strengths $C_{6,mm'}$ as functions of $\xi$. In Figs.~\ref{effpot}(b)-(c), we compare the effective potential $V_{\mathrm{eff}%
}({\boldsymbol{r}})$ (dashed lines) with the adiabatic potential $V^{%
\mathrm{(ad)}}({\boldsymbol{r}})$ (solid lines) along the $y$- and $x$-
directions for various values of $\xi$. The attractive
interaction along the $x$-direction is suppressed and along the $y$%
-direction is enhanced for $\xi>0$. The analytical expression in Eq.~(%
\ref{Veff}) provides a highly accurate form for the molecular interaction,
which is crucial for investigating many-body physics. 

\begin{figure}[tbp]
\includegraphics[width=0.8\linewidth]{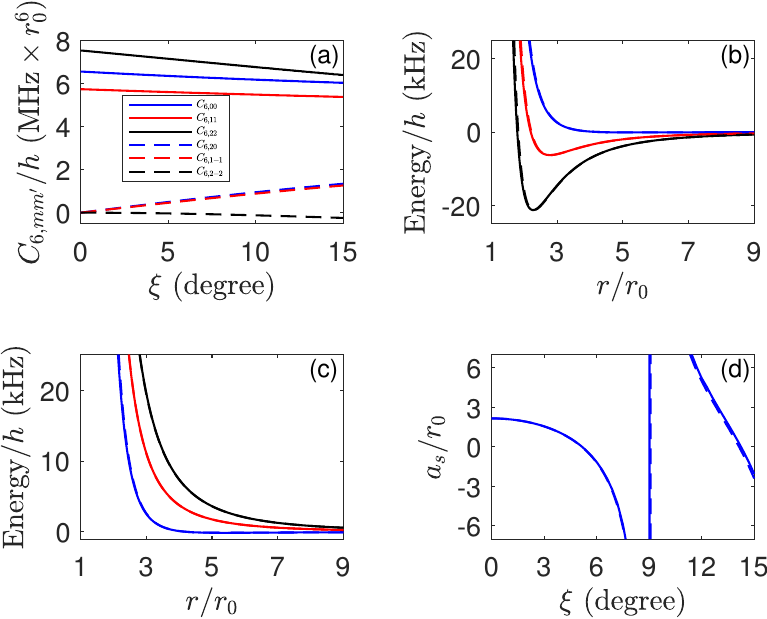}
\caption{(a) Elements of the matrix $C_6$ as functions of $\xi$. 
(b)-(c): Adiabatic and effective potentials for \(\xi = 0^{\circ}\) (blue), \(3^{\circ}\) (red), and \(6^{\circ}\) (black). Solid curves represent the adiabatic potentials, while dashed curves show the corresponding effective potentials. (b) Potentials along the \(y\)-direction. (c) Potentials along the \(x\)-direction.
(d) Excellent agreement between the \(s\)-wave scattering lengths \(a_s\) obtained from the effective potential and from the full coupled-channel calculation.
The microwave parameters are: \(\delta_\sigma/(2\pi) = -8\,\textrm{MHz}\), \(\Omega_\sigma/(2\pi) = 7.9\,\textrm{MHz}\), \(\delta_\pi/(2\pi) = -10\,\textrm{MHz}\), and \(\Omega_\pi/(2\pi) = 6.5\,\textrm{MHz}\).
}
\label{effpot}
\end{figure}

\section{Multichannel scattering computation in Floquet space}

\label{appflosc}

In this Appendix, we turn to formulate the multi-channel scatterings for
Floquet-Bloch states. To solve Eq.~\eqref{FE}, we expand the eigenstate
wavefunction $\psi _{n,\nu }({\boldsymbol{r}})=\langle \nu |\psi _{n}({
\boldsymbol{r}})\rangle $ in the partial-wave basis as
\begin{equation}
\psi _{n,\nu }({\boldsymbol{r}})=\frac{1}{r}\sum_{lm}\phi _{n\nu
lm}(r)Y_{lm}(\hat{\boldsymbol{r}}),  \label{expansion}
\end{equation}
where $\phi _{n\nu lm}$ are the radial wavefunctions.

To proceed further, it is convenient to introduce the column vector ${
\boldsymbol{\Phi }}({\boldsymbol{r}})$ formed by the elements $\phi _{n\nu
lm}(r)$. The Schr\"{o}dinger equation for the radial wavefunction can then
be written as
\begin{equation}
\partial _{r}^{2}{\boldsymbol{\Phi }}(r)+\mathbf{W}(r){\boldsymbol{\Phi }}
(r)=0,  \label{SE2}
\end{equation}
where the matrix $\mathbf{W}$ is defined by the elements 
\begin{align}
W_{n\nu lm,n^{\prime }\nu ^{\prime }l^{\prime }m^{\prime }}(r)& =\left[
k_{n\nu }^{2}-\frac{l(l+1)}{r^{2}}\right] \delta _{nn^{\prime }}\delta _{\nu
\nu ^{\prime }}\delta _{ll^{\prime }}\delta _{mm^{\prime }}-M\sum_{s=-3}^{3}\delta _{n+s,n^{\prime }}V_{\nu lm,\nu ^{\prime
}l^{\prime }m^{\prime }}^{(s)}(r)
\end{align}
with $k_{n\nu }=\left[ M\left( \varepsilon -\mathcal{E}_{\nu \nu }^{(\infty
)}+n\omega \right) \right] ^{1/2}$ being the incident momentum with respect
to the $n$th Floquet sector and the $\nu $th dressed channel and 
\begin{equation}
V_{\nu lm,\nu ^{\prime }l^{\prime }m^{\prime }}^{(s)}(r)=\int d\hat{
\boldsymbol{r}}Y_{lm}^{\ast }(\hat{\boldsymbol{r}})[\mathcal{V}_{s}({
\boldsymbol{r}})]_{\nu \nu ^{\prime }}Y_{l^{\prime }m^{\prime }}(\hat{
\boldsymbol{r}})
\end{equation}
can be evaluated analytically.

To solve the multi-channel scattering problem, we numerically evolve Eq.~
\eqref{SE2} from a ultraviolet cutoff $r_{\mathrm{UV}}$ to a sufficiently
large value $r_{\infty }$ using the log-derivative propagator method. Here,
we impose a capture boundary condition~\cite{Karman2018} at $r_{\mathrm{UV}}$
. It turns out that the choice of $r_{\mathrm{UV}}$ does not affect the
result if it is deep inside the shielding core~\cite
{Deng2023,Deng2024,Deng2025}. We remark that to account for the short-range
effects in scatterings, we also include, in numerical calculations, the
universal van der Waals interaction through the replacement $\mathcal{V}_{0}(
{\boldsymbol{r}})\rightarrow \mathcal{V}_{0}({\boldsymbol{r}})-C_{\mathrm{vdW
}}/r^{6}$, where $C_{\mathrm{vdW}}$ is the strength of the universal van der
Waals interaction~\cite{Julienne2010}. Since $C_{\mathrm{vdW}}$ is generally
much smaller than the microwave shielding strength $C_{6}$~\cite{Lepers2013}
, the $C_{\mathrm{vdW}}$ term only takes effect at short distance.

Then to obtain the scattering matrix, we compare $\phi _{n\nu
lm}(r_{\infty })$ with the asymptotic boundary condition
\begin{align}
\phi _{n\nu lm}(r)& \overset{r\rightarrow \infty }{\longrightarrow }
k_{n_{0}\nu _{0}}^{-1/2}\hat{\jmath}_{l}(k_{n_{0}\nu _{0}}r)\delta
_{nn_{0}}\delta _{\nu \nu _{0}}\delta _{ll_{0}}\delta _{mm_{0}}+k_{n\nu }^{-1/2}\hat{n}_{l}(k_{n\nu }r)K_{n\nu
lm,n_{0}\nu _{0}l_{0}m_{0}},
\end{align}
where $\hat{\jmath}_{l}(z)$ and $\hat{n}_{l}(z)$ are the Riccati-Bessel
functions, and $K_{n\nu lm,n_{0}\nu _{0}l_{0}m_{0}}$ are elements of the $K$
matrix, corresponding to the scattering from the incident channel $(n_{0}\nu
_{0}l_{0}m_{0})$ to the outgoing one $(n\nu lm)$. Moreover, $k_{n_{0}\nu
_{0}}$ and $k_{n\nu }$ are the relative momenta for the incident and
outgoing channels, respectively. In numerical calculations, we introduce a
truncation $n_{\mathrm{cut}}$ for the Floquet Hamiltonian such that $|n|\leq
n_{\mathrm{cut}}$. Practically, it is found that, for control parameters
covered in this work, the scattering solutions converge when $n_{\mathrm{cut}
}=5$.

To proceed further, we denote the $K$ matrix as ${\mathbf{K}}$, from which
one can obtain the scattering matrix $\mathbf{S}=\left( 1+i\mathbf{K}\right)
\left( 1-i\mathbf{K}\right) ^{-1}$. Now, the total elastic cross section for
the incident channel $(n_{0}\nu _{0})$ is
\begin{equation}
\sigma _{n_{0}\nu _{0}}^{(\mathrm{el})}=\frac{2\pi }{k_{n_{0}\nu _{0}}^{2}}
\sum_{lml_{0}m_{0}}\left\vert \delta _{ll_{0}}\delta _{mm_{0}}-S_{n_{0}\nu
_{0}lm,n_{0}\nu _{0}l_{0}m_{0}}\right\vert ^{2},
\end{equation}
where $S_{n_{0}\nu _{0}lm,n_{0}\nu _{0}l_{0}m_{0}}$ are the elements of the
scattering matrix. $l$ and $l_{0}$ are even (odd) for bosons (fermions). For
elastic scattering, the outgoing channel of the molecules is the same as
incident channel $(n_{0}\nu _{0})$ and the total kinetic energy of the
molecules is thus conserved. Next, the total inelastic cross section can be
calculated by subtracting the total elastic cross section from the total
cross section, i.e., 
\begin{equation}
\sigma _{n_{0}\nu _{0}}^{(\mathrm{inel})}=\frac{2\pi }{k_{n_{0}\nu _{0}}^{2}}
\sum_{lml_{0}m_{0}}\left( \delta _{ll_{0}}\delta _{mm_{0}}-|S_{n_{0}\nu
_{0}lm,n_{0}\nu _{0}l_{0}m_{0}}|^{2}\right) .  \label{eqinel}
\end{equation}
It is instructive to distinguish, depending on whether the total energy of
the colliding molecules is conserved, the degenerate and nondegenerate
inelastic scatterings which are disguised in Eq.~\eqref{eqinel}.
Specifically, for degenerate inelastic scatterings, the outgoing molecules
remain in the same Floquet sector $n_{0}$ but transit to the lower
dressed-state channel $\nu $ ($>\nu _{0}$). Thus the total energy of the
colliding molecules is conserved. While for nondegenerate inelastic
scatterings, the outgoing molecules transit to a distinct Floquet sector $n$
($\neq n_{0}$) by absorbing or emitting microwave photons and thus the total
energy of the colliding molecules is not conserved. These the energy
exchange processes mediated by the absorption and emission of dual
microwaves lead to inelastic scattering and heating.

The experimentally more relevant quantities are the elastic and inelastic
scattering rates, i.e., $\beta _{01}^{(\mathrm{el})}=v_{01}\sigma _{01}^{(
\mathrm{el})}$ and $\beta _{01}^{(\mathrm{inel})}=v_{01}\sigma _{01}^{(
\mathrm{inel})}$, where $v_{01}=2\hbar k_{01}/M$ is the relative velocity.
In addition, the ratio of the elastic to inelastic scattering rates, $\gamma
=\beta _{01}^{(\mathrm{el})}/\beta _{01}^{(\mathrm{inel})}$, is of
particular importance for characterizing the efficiency of the evaporative
cooling. Finally, from the $K$ matrix, we may compute for small $k_{01}$ the
scattering length matrix according to $\mathbf{a}=-\mathbf{K}/k_{01}$, whose
element $a_{n\nu lm,n_{0}\nu _{0}l_{0}m_{0}}$ is the scattering length from
the incident channel $(n_{0}\nu _{0}l_{0}m_{0})$ to the outgoing channel $
(n\nu lm)$. In particular, the $s$-wave scattering length for MSPMs is $
a_{s}=a_{0100,0100}$.

To validate the effective potential, we also solve the Schr\"{o}dinger
equation
\begin{equation}
H_{\mathrm{eff}}\psi ({\boldsymbol{r}})=\frac{\hbar ^{2}k_{01}^{2}}{M}\psi ({
\boldsymbol{r}})
\end{equation}
using log-derivative propagator method. Here, $k_{01}$ is the incident
momentum, and the single-channel model for the relative motion of two MSPMs
is then governed by the Hamiltonian
\begin{equation}
H_{\mathrm{eff}}=\frac{{\boldsymbol{p}}^{2}}{M}+V_{\mathrm{eff}}({
\boldsymbol{r}}).
\end{equation}
Making use of the partial-wave expansion, $\psi ({\boldsymbol{r}}
)=\sum_{lm}r^{-1}\phi _{lm}(r)Y_{lm}(\hat{\boldsymbol{r}})$, the radial wave
functions satisfy
\begin{align}
-\frac{1}{M}& \left[ \frac{d^{2}}{dr^{2}}-\frac{l(l+1)}{r^{2}}\right] \phi
_{lm}(r)+\sum_{l^{\prime }m^{\prime }}[V_{\mathrm{eff}}(r)]_{lm,l^{\prime
}m^{\prime }}\phi _{l^{\prime }m^{\prime }}(r)=\frac{k_{01}^{2}}{M}\phi
_{lm}(r),  \label{radiawfeff}
\end{align}
where the interaction matrix elements
\begin{equation}
\lbrack V_{\mathrm{eff}}(r)]_{lm,l^{\prime }m^{\prime }}=\int d\hat{
\boldsymbol{r}}Y_{lm}^{\ast }(\hat{\boldsymbol{r}})V_{\mathrm{eff}}({
\boldsymbol{r}})Y_{l^{\prime }m^{\prime }}(\hat{\boldsymbol{r}}).
\label{mateleveff}
\end{equation}
In analogy to the multichannel case, we numerically solve Eqs.~
\eqref{radiawfeff} and compute the low-energy scattering length $
a_{lm,l_{0}m_{0}}$ for the scattering from incident channel $(l_{0}m_{0})$
to the outgoing one $(lm)$. Particularly, we focus on the $s$-wave
scattering length $a_{s}=a_{00,00}$.

In Fig.~%
\ref{effpot}d, we compare the $s$-wave scattering lengths $a_{0100,0100}$ and $a_{00,00}$ obtained
from the full-channel calculation and the effective potential (%
\ref{Veff}) as a function of the elliptic angle $\xi>0$. As $\xi $
increases from zero, the positive $C_{3,2}$ deepens the potential along
the $y$-direction. When $\xi $ reaches $\xi _{\mathrm{res}}=9.2^{\circ }$,
the ebahnced attractive interaction beocmes strong enough to induce a shape
resonance. Beyond this angle, a bound state emerges in the potential valley
[see Fig.~\ref{effpot}b]. Notably, the two approaches yield
quantitatively consistent results away from resonance. Even near the
resonance, excellent agreement is maintained, with the resonance position
differing by less than $0.4^{\circ }$. These results confirm the reliability
of the effective potential $V_{\mathrm{eff}}({\boldsymbol{r}})$ in capturing molecular collision dynamics under an elliptically polarized microwave field.

\section{Superfluid fractions}
\label{SF}
\label{SF} 
\begin{figure*}[t]
\includegraphics[width=0.8\linewidth]{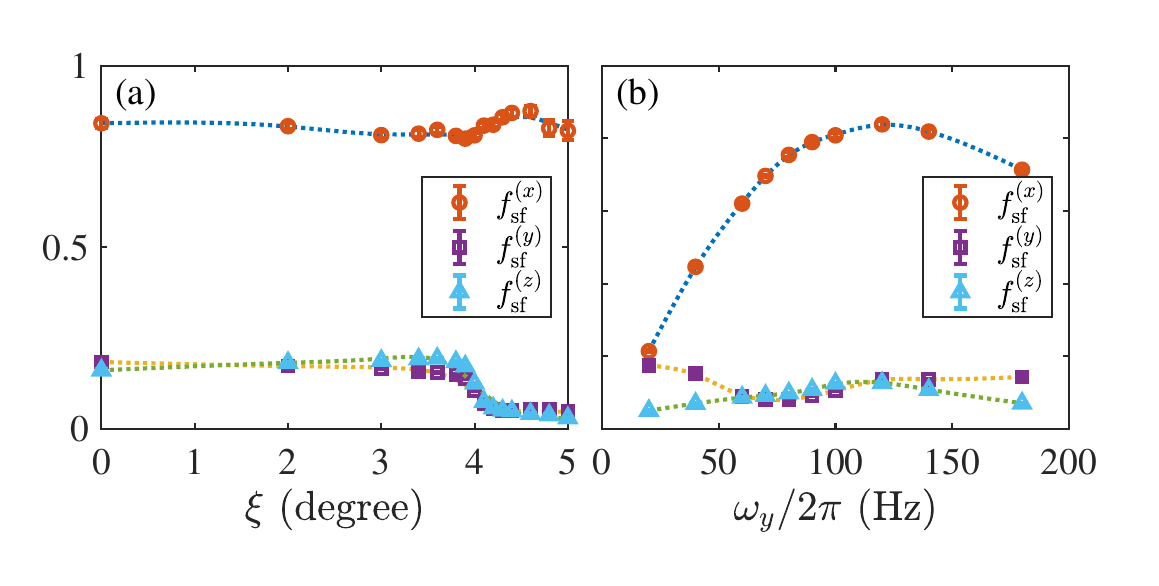}
\caption{Superfluid fractions are averaged over 40-64 random seeds for a fixed trap frequency $\omega_y=2\pi\times100$Hz (a) and for a fixed ellipticity $\xi=4^\circ$ (b).}
\label{superfrac}
\end{figure*}

In Fig.~\ref{superfrac}, we show the superfluid fractions $f_{\rm sf}^{(x,y,z)}$ by changing the ellipticity and the trap frequency $\omega_y$. 

Figure~\ref{superfrac}(a) shows the superfluid fraction for a fixed trap frequency $\omega_y=2\pi\times100$Hz. Among the three directions, the superfluid fraction $f_{\rm sf}^{(x)}$ is dominant, reflecting higher stiffness
corresponding to a larger critical rotational frequency along the $x$ axis. The anisotropy of $f_{\rm sf}^{(x,y,z)}$ arises from the interplay between the trap geometry and DDI. For $\xi = 3.9^{\circ}$ and $4.0^{\circ}$, the persistence of a dominant $%
f_{s,x}$ implies that coherence is established within each droplet of the
array. Meanwhile, the finite values of $f_{\rm sf}^{(y,z)}$ suggest phase
coherence between distinct droplets, a hallmark of supersolidity. As $\xi$ increases
beyond $4.2^{\circ}$, $f_{\rm sf}^{(y,z)}$ drop
significantly, indicating the breakdown of long-range phase coherence
between droplets in the array along the $x$ direction. For even larger elliptic angles, such as $\xi = 4.6^{\circ}$, the
strong long-range attraction along the $y$ axis collapses the system into a
single, high-density droplet. In this regime, the narrow size of the droplet suppresses the exchange of world lines in both the \( x\text{-}z \) and \( x\text{-}y \) planes, and inhibits phase coherence along the \( x \)-direction, as coherence typically builds up over several \( r_0 \) due to the strong shielding potential. As a result, both $f_{\rm sf}^{(y,z)}$ are reduced.  

Figure~\ref{superfrac}(b) shows the superfluid fraction for a fixed elliptic angle $\xi=4^\circ$. For a small trap frequency $\omega_{y} =2\pi\times 20\,\text{Hz}$, the superfluid fraction $f_{\rm sf}^{(x)}$ is much lower than that in Fig.~\ref{superfrac}(a), as a result of a larger size in the \( y \)-\( z \) plane [cf. Figs. 4(a1) in the main text]~\cite{RevModPhys.81.647}. As the trap is compressed (e.g., $%
\omega_{y}/(2\pi) = 60, 80, 100\,\text{Hz}$), the system crystallizes along
the $x $-direction, forming a droplet array with lower density. The reduced size in the \( y \)-direction enhances superfluid fraction $f_{\rm sf}^{(x)}$, while the nonvanishing components $f_{\rm sf}^{(y,z)}$ indicate coherence between different droplets and the emergence of a supersolid phase. This occurs because the droplets begin to merge, and the increased interparticle distance along the \( x \)-direction reduces the influence of short-range repulsion. With further increase in trap frequency beyond $120$Hz, each droplet elongates along the $x 
$-direction. Eventually, the droplets merge together, forming a continuous gas
extended along the $x$-direction. In this regime, most molecules are redistributed from the
attractive ($y $-) direction to the repulsive ($z $- and $x $-) directions. The narrow size along the y-direction makes it difficult to exchange worldlines in the $y$-$z$ plane~\cite{RevModPhys.67.279}, thereby reducing $f_{\rm sf}^{(x)}$.

\end{document}